\documentclass{ws-procs9x6}

\usepackage{graphicx}

\newcommand{\met}{$E\!\!\!\!/_{T}$}
\newcommand{\lsp}{\mbox{$\tilde{\chi}_{1}^{0}$}}
\newcommand{\chargino}{\mbox{$\tilde{\chi}_{1}^{\pm}$}}
\newcommand{\ipb}{\mbox{${\rm pb}^{-1}$}}

\begin{document}
\title{New Physics Searches with Photons in 
  CDF\footnote{\uppercase{T}alk presented by
  \uppercase{S.W.~L}ee at {\it \uppercase{SUSY} 2003:
  \uppercase{S}upersymmetry in the \uppercase{D}esert}\/, held at the
  \uppercase{U}niversity of \uppercase{A}rizona, \uppercase{T}ucson,
  \uppercase{AZ}, \uppercase{J}une 5-10, 2003. \uppercase{T}o appear
  in the \uppercase{P}roceedings.}\\[-0.5cm]
}

\author{Sungwon Lee \lowercase{for the} CDF~Collaboration}

\address{Department of Physics \\
Texas A\&M University \\
College Station,  TX77843, USA\\ 
E-mail: slee@fnal.gov}

%


\maketitle

\abstracts{\vspace*{-0.3cm}
A brief review of searches for physics beyond the Standard Model  with photons
using the CDF detector at the Tevatron is given here. These include searches
for supersymmetry, extra dimensions, excited electrons and W/Z$+\gamma$
production, as well as anomalous photon production. Recent results from CDF Run
II experiment is presented, but some results from Run I are also reviewed. 
}

\vspace*{-0.5cm}
\section{Introduction}
There are a large number of important and well-motivated theoretical models
which make a strong case for looking for new physics in events with a photon in
the final state. These theories include Supersymmetry~(SUSY), Extra
Dimensions~(ED), Grand Unified Theories, Composite models, Anomalous couplings
and Higgs models. In experimental side CDF detected the $ee\gamma\gamma$\met\
candidate event which has never been explained by conventional
physics~\cite{eeggmet}.   

Besides the specific theoretical models, searching for new physics with photons
has several advantages. For example, the photon is one of the three $SU(2)
\times U(1)$ gauge bosons and as such is likely to be a good probe of new
interactions since it couple to any new gauge sector. Final-state photons have
additional distinct detection advantages over W$^\pm$ or Z$^{0}$ bosons since
they do not decay. Thus they do not suffer a  sensitivity loss from branching
ratios and momentum sharing between the decay products. There are very few
Standard Model~(SM) backgrounds which produce photons, allowing a fairly clean
signature. 

In this article we summarize the current CDF experimental results of searches
for new physics in final states containing energetic photons at the Tevatron.

\begin{figure}[ht]
\centerline{\epsfxsize=0.75\textwidth 
  \epsfbox{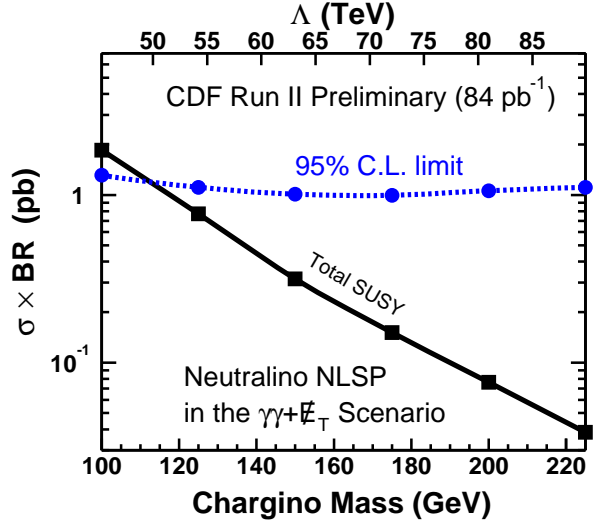}}
\vspace{-0.2cm}   
\caption{\label{fig:limit}
The $95\,\%$ C.L. cross section limits from $\gamma\gamma+$\met\ analysis,
without background subtraction. Also shown is the total SUSY cross section for
the neutralino NLSP as a function of M$_{\chargino}$ and $\Lambda$.}
\end{figure}

\vspace*{-0.15cm}
\section{Search for Supersymmetry in $\gamma\gamma+$\met\ Events}
Among various SUSY models, two SUSY breaking mechanisms are interested, which
predict photons in the final state. Supergravity models can produce events
which decay down to the second lightest neutralino via a loop into the lightest
neutralino~(\lsp) and a photon, where the \lsp\ is the lightest supersymmetric
particle~(LSP).  Gauge-Mediated SUSY Breaking models~(GMSB) with  the \lsp\
decaying into a photon and gravitino can produce a final state of two photons
and large missing transverse energy~(\met). \met\ is often used as a pointer to
possible SUSY signals because indicates the escape of a non-interacting SUSY
particle from the detector.   The LSP signals are of particular interest as
they provide  a natural explanation for the dark matter.

CDF has searched for $\gamma\gamma$+\met\ final state within GMSB scenario
using 84 \ipb\  of Run II data. Events are selected as having two photon
candidates with $E_T$$>$ 13  GeV in the central. We expect a total of 
$0.77^{+0.40}_{-0.21}$ events with \met\ $>$ 25 GeV and two events pass all
requirements in the data.  The lower mass limit on the lightest chargino
derived from this analysis is  M$_{\chargino} >$ 113 GeV at the $95\,\%$
C.L.~lower limit (see Figure 1). Updated analysis is now underway using 200
\ipb\  data and new results are starting to appear. 

\vspace*{-0.15cm}
\section{Searches for Anomalous $\gamma\gamma$ production}

CDF has searched for new physics in the Run II diphoton sample. Two isolated
photons, each with $E_T >$ 25 GeV, are required in the analysis. The main
background comes from jets which fake photons. No evidence for new physics was
found in this sample. CDF has also performed a search in the diphoton sample
for events with an additional lepton~($e$ or $\mu$) and estimate the
backgrounds to each of these measurements. All data are well described by the
SM expectations.

\vspace*{-0.15cm}
\section{Searches for Anomalous $l\gamma$ production}

CDF has performed a model independent search for anomalous production of events
with a high $E_T$ photon and a lepton~($e$ or $\mu$) in the final state using
Run I data~\cite{lg}.  Several final states were defined, based on the presence
of leptons, photons, and \met. All data sets are consistent with the SM
expectations with a possible exception of $l\gamma +$\met\  were 7.6 $\pm$ 0.7
events are expected while 16 events are observed. CDF experiment continue to
search the anomalous  $l\gamma$ production using Run II data for hints of new
physics. 

\vspace*{-0.15cm}
\section{Searches for Extra Dimensions}
Recent theories postulate the existence of new space-time dimensions. Such
extra dimensions might be found by studying the emission of the electromagnetic
radiation in Graviton~($G$) into the EDs, together with single photon or
diphoton emitted into the normal dimensions.

CDF has searched for direct $G$ production in the $\gamma G$ final state using
Run II data. The analysis required a photon with $E_T$$>$ 47 GeV, \met\ $>$ 42
GeV,  and no jets with $E_T$$>$ 10 GeV in the event. The main backgrounds are
from cosmic rays and Z $\rightarrow \nu\bar{\nu}\gamma$. Total 19.8 $\pm$ 2.3
events are expected from the backgrounds, and 18 events are observed. No
deviation from the SM expectation is observed, and the limit is derived as a
ratio to the expected background from the irreducible SM process $q\bar{q}
\rightarrow$ Z$\gamma \rightarrow \nu\bar{\nu}\gamma$. The limit is 2 times the
expected Z$\gamma$ signal. This is a significant improvement over the Run I,
which obtained a limit 3.1 times the Z$\gamma$ signal.

Another way to search for the ED is to look for $G$ exchange processes in the
diphoton final state and looking for excess in the invariant mass distribution.
In Run I CDF found no evidence of a signal, and the $95\,\%$ C.L.~lower limit
on the effective Plank scale in the ED, M$_{S}$, were set at 989~(853) GeV for
the Hewett convention, $\Lambda_{\rm Hewett}$ = -1~(+1). A search for a
Randall-Sundrum graviton in the diphoton decay mode is now underway using Run
II data and preliminary results are
starting to appear.

\vspace*{-0.15cm}
\section{Searches for excited electrons with $e\gamma+e$ Events}
CDF uses 72 \ipb\  of Run II data to search for the production of excited
electrons~($e^{*}$) using the reaction $p\bar{p}\rightarrow e^{*}+e\rightarrow
e\gamma+e$. This is a signature-based search for a central $ee\gamma$ final
state with a resonance in the $e\gamma$ channel.  This analysis required two
high $p_T$ electrons with an additional photon in the final state.   CDF
observed no candidate events in the data after making all selections, and set
$95\,\%$ C.L. limit on the production cross section times BR($e^{*}\rightarrow
e\gamma$), and on the $e^{*}$ mass for various choices of the compositeness
scale $\lambda$ in the $e^{*}$ model. For M$_{e^{*}}$=$\lambda$, the mass limit
is 785 GeV.

\vspace*{-0.15cm}
\section{Searches for W/Z$+\gamma$ production}
The associated production of a vector boson and a photon is an ideal test of
the triple gauge couplings which are precisely predicted by the SM. Any
deviation from the SM could indicate new physics. Several measurement are
preformed by the CDF using 73 \ipb\  of Run II data. All available
lepton triggers are used to select W and Z candidates inclusively and the
additional photon is then selected. 

The cross section for W$\gamma$/Z$\gamma$ production are measured for $\Delta
R_{l\gamma}$$>$0.7 and $E_T^{\gamma}$$>$7~GeV and the kinematic distributions
are compared to the SM prediction. The W$\gamma$ cross section is measured to
be 18.2 $\pm$ 2.9(stat) $\pm$ 2.3(sys) $\pm$ 1.1(lumi)~pb compared to the SM
expectation of 18.7 $\pm$ 1.3(theory)~pb. The Z$\gamma$ cross section is found
to be 5.8 $\pm$ 1.3(stat) $\pm$ 0.7(sys) $\pm$ 0.3(lumi)~pb compared to the
theoretical prediction of 5.3 $\pm$ 0.4(theory)~pb. The results are in
excellent  agreement with the SM expectations.

\vspace*{-0.15cm}
\section{Conclusion}
Since photon is a clean and well measured electromagnetic object, new physics
searches with photons are particularly interesting. CDF experiment is taking
data actively since 2001 and larger samples are being collected for new physics
searches based on photon signature. High luminosity data with photon will
provide a good opportunity for new physics discoveries, and will give
experimental guidance to a better theoretical modeling of new physics
production with photon in the final states.


\vspace*{-0.15cm}
\begin{thebibliography}{0}

\bibitem{eeggmet} CDF Collaboration, 
Phys. Rev. Lett. {\bf 81}, 1791 (1998).

\bibitem{lg} CDF Collaboration, 
Phys. Rev. Lett. {\bf 89}, 041802 (2002).

\end{thebibliography}
\end{document}